\documentstyle[amstex,amssymb,12pt]{article}

\def\ra{\rightarrow}

\def\nn{\nonumber}

\font\fortssbx=cmssbx10 scaled \magstep3

\begin{document}

\begin{titlepage}

\begin{center}
{\fortssbx University of Tennessee, Knoxville}
\end{center}

\vspace{ .10in}

\begin{flushright}
{\bf UTHEP-95-0501}\\
{May 1995}
\end{flushright}

\vspace{ .50in}

\begin{center}
{\LARGE\bf Multiple Gluon Effects in $q + \bar q \ra
t + \bar t + X$ at FNAL Energies: Semi-Analytical Results
\footnote{Work supported in part by the US DoE under grant
DE-FG05-91ER40627 and by Polish Government grant KBN 2P30225206}}
\\[.4in]
D. DeLaney, S. Jadach \footnote{Permanent Address:
Institute of Nuclear Physics, ul. Kawiory 26a, Cracow, Poland},
S. M. Kim, Ch. Shio, G. Siopsis, and B. F. L. Ward
\\[.2in]
{\it Department of Physics, University of Tennessee \\
Knoxville, Tennessee 37996-1200, U.S.A.}
\end{center}

\vspace{ .3in}

\begin{abstract}
{
We apply our Yennie-Frautschi-Suura exponentiated
cross section formulas for
the parton processes
$q + {^(} \bar q {^)}{^\prime} \ra
 q{^\prime}{^\prime} +
 {^(} \bar q {^)}{^\prime}{^\prime}{^\prime} + n(G)$
to the process
$q + \bar q \ra t + \bar t + n(G)$ at FNAL energies,
where G is a QCD gluon.  We use semi-analytical methods to compute the
ratio $r_{exp}=\sigma_{exp}/\sigma_B$, where $\sigma_{exp}$ is our soft gluon
YFS exponentiated cross section and $\sigma_B$ is the Born cross section.
For $m_t= 0.176(0.199)$TeV, we get $r_{exp}=1.65(1.48)$, respectively,
for $q=u$ for example.
These results are not inconsistent with the recent observations
by CDF and D0.
}
\end{abstract}

\end{titlepage}


In the recent observations by CDF~\cite{CDF} and D0~\cite{D0}
of strong evidence for the long awaited top quark,
at the respective masses
$m_t = 0.176 \pm 0.008 \pm .01$ TeV,
$m_t = 0.199^{+0.019}_{-0.02} \pm 0.02$ TeV,
there is the further result that
the respective observed cross sections,
$\sigma (t\bar t)_{\rm obs} = 6.8^{+3.6}_{-2.4}$ pb,
$6.4 \pm 2.2$ pb, are
on average a factor $\sim 2$ larger than the theoretically predicted
cross section, the $\sigma(t\bar t)_{\rm th}$, from Ref.~\cite{res}.
Of course, the three sets of cross sections are within $3\sigma$ of one
another so that one can only say that there is
a suggestion in the data that
certain higher order corrections have not been
taken into account in the
predictions in Ref.~\cite{res}.
We use a semi-analytical representation
of the recently introduced~\cite{five}
Yennie-Frautschi-Suura (YFS)~\cite{YFS} soft gluon exponentiated
parton level top production cross section of five of us
(D. D., S. J., Ch. S., G. S., and B.F.L. W.)
to compute the effects of multiple soft gluon radiation on
$\sigma(t\bar t)_{th}$ at the parton level.
A detailed analysis of the corresponding
effects at the level of events, both at the parton level and at
the hadron-hadron scattering level, is in progress and will
appear elsewhere~\cite{elsewhere}.

We need to stress as well that the predictions in Ref.~\cite{res}
involve an extra non-perturbative soft parameter
(in addition to the usual factorization and renormalization scales)
in the soft gluon resummation and
this parameter introduces an uncertainty in the respective cross section
normalization. Thus, independent of the suggestion that
the ratio $      \sigma(t\bar t)_{\rm obs}/\sigma(t\bar t)_{\rm th}$
may be significantly different from 1,
one would like to have a unique prediction for the soft gluon effects in
$\sigma(t\bar t)_{\rm th}$.
We will see below that our prediction for the soft gluon effects in
$\sigma(t\bar t)_{\rm th}$ does not suffer from such unknown
non-perturbative parameters as those
involved in Ref.~\cite{res} -- we get a unique prediction for the effects
of soft gluons in $\sigma (q\bar q \ra t\bar t)$ and
thereby for those in  $\sigma(p\bar p\ra t\bar t)$~\cite{elsewhere}.

Specifically, from Ref.~\cite{five},
we have a representation of the basic
soft gluon YFS exponentiated cross section for
$q(Q_1)+\bar q(P_1) \ra t(Q_2) \bar t(P_2) + n(G)(k_1,\cdots,k_n)$ in an
obvious kinematics notation as
(we note that at FNAL energies this sub-process
actually dominates $p\bar p \ra t\bar t +X$)
\begin{eqnarray}
d\sigma_{\rm exp}
&=& {\rm exp} \Big[ {\rm SUM_{IR} (QCD)} \Big]
\sum_{n=0}^{\infty} \int \prod_{j=1}^n {d^3k_j\over k_j}
\int {d^4y \over (2\pi)^4}
e^{iy(P_1+Q_1-P_2-Q_2-\sum k_j)
+D_{\rm QCD}}\nn\\
&& \qquad\qquad\qquad \times \bar \beta_n (k_1,\cdot\cdot\cdot,k_n)
{d^3Q_2 \over Q_2^0} {d^3P_2 \over P_2^0}
\label{csexp}
\end{eqnarray}
with~\cite{five}
\begin{eqnarray}
D_{\rm QCD}&=&\int {d^3k \over k} \tilde S_{\rm QCD} (k)
  ( e^{-iy\cdot k} - \theta (K_{\rm max} - k) ), \\
{\rm SUM_{IR} (QCD)} &=& 2\alpha_s Re B_{\rm QCD}
+2\alpha_s\tilde B_{\rm QCD} (K_{\rm max}),
\end{eqnarray}
for ($m_G$ is our gluon infrared regulator mass)
\begin{equation}
{2\alpha_s}\tilde{B}_{\rm QCD} = \int^{k\leq K_{\rm max}}
{d^3 k \over (k^2+m_G^2)^{1/2}} \tilde S_{\rm QCD} (k);
\end{equation}
and here~\cite{five}
the hard gluon residuals, $\bar\beta_n$, are defined in complete
analogy with the YFS hard photon residuals in QED; in particluar, to the
order to which we shall work in this paper, we will only need the
residual $\bar\beta_0$ which is just $2d\sigma_B/d\Omega_t$
where $\sigma_B$ is the respective Born cross section and $d\Omega_t$
is the respective elemental solid angle of a detected top quark, for
example. Explicit
formulas for the YFS QCD infrared functions
$B_{QCD}$ and $\tilde S_{QCD}$
can be found in Refs.~\cite{five,rgYFS} and
references therein, for example.
We stress that cross section in eq.~(\ref{csexp})
is independent of the dummy parameter $K_{max}$.

Let us now emphasize that by working to the $\bar\beta_0$ level in
eq.~(\ref{csexp}) we obtain an infinite order summation of the soft gluon
effects which is exact in the soft gluon regime. It has been shown in
Refs.~\cite{res} that the hard gluon corrections at ${\cal O}(\alpha_s)$
relative to the Born cross section are, after removable of the genuine
pure collinear singularities of the massless limit, in fact small.
Thus, the level of our approximations in this paper is well matched
to the realistic parton-parton and hadron-hadron $t\bar t$ production
processes at FNAL and our results for the effects of soft gluons
should transcribe directly to these processes.

Specifically, if we retain the $\bar\beta_0$ term in eq.(\ref{csexp}),
we get the analytical result, the QED analogue
of which is already presented in Ref.~\cite{rgYFS,jrww},
\begin{eqnarray}
{{d\sigma(t\bar t)_{exp}}\over{dvd\Omega_t}} &=&
                                {{\gamma F_{YFS}(\gamma)} \over v}
   {{d\sigma(t\bar t)_B(v)}\over {d\Omega_t}}
{e^{\{2\alpha_s\{ ReB_{QCD}  + \tilde B_{QCD}(v)\}\}}},
\label{sigex}
\end{eqnarray}
where $v=(s-s')/s$,
 $F_{YFS}(\gamma)$ is the famous YFS function $e^{-C\gamma}/\Gamma(1+
\gamma)$, and the radiation probability unit $\gamma$ is defined by
\begin{eqnarray}
\gamma  &=&k^2\int d\Omega_k \tilde S_{QCD}(k) \nonumber \\
        &=&{\alpha_s\over \pi}\{ 2C_F(ln{s\over m_q^2}-1)+2(C_F-C_A/2)
        ln{s\over m_q^2}\},~~initial~state~radiation
\label{gma}
\end{eqnarray}
where $C_{F(A)}$ is the quadratic Casimir invariant for the
vector(adjoined) representation of SU(3) color respectively
and we show the dominant initial state radiation probability unit
only for explicit illustration~\cite{elsewhere}.
Here, $C=0.5772156...$  is Euler's constant, $s=(Q_1+P_1)^2$,
$s'=(Q_2+P_2)^2$, and $d\Omega_k$ is the elemental solid angle
for a gluon of 4-momentum $k$.
The result (\ref{sigex}) is the fundamental result of this paper.
We emphasize that it is a rigorous consequence of the
Feynman-Schwinger-Tomonaga
series for QCD and that it takes into account the effects of soft
gluon radiation to all orders in $\alpha_s$ without the introduction
of arbitrary soft scales.

Upon integrating (\ref{sigex}) numerically over $v$,
from $0$ to $v_{max}=1-4m_t^2/s$, for the FNAL beam energy we get,
for the typical $\alpha_s\cong .082$, the basic results
for $u\bar u$ annihilation using
$m_u($1GeV$)\cong 5.1$MeV~\cite{leug},
where $\gamma\simeq 1.62$,
\begin{eqnarray}   r_{exp}\equiv
  \sigma(t\bar t)_{exp}/\sigma(t\bar t)_B
&=&
 {1\over \sigma(t\bar t)_B}
 \int\limits_{0}^{v_{max}}dv{d\sigma(t\bar t)_{exp}(v)\over{dv}}
\nonumber \\
 &=& \gamma F_{YFS}(\gamma)
    \exp\left\{ {\alpha_s\over\pi}[(2C_F-{1\over 2}C_A)
         ({1\over 2}ln{s\over m_u^2}+{\pi^2\over 3})-C_F] \right\}
\nonumber \\
    &\times&{{\int_{0}^{v_{max}}dvv^{\gamma-1}  (v_{max}-v)^{1\over 2}
   ({3\over 2}-v-{1\over 2}v_{max})}\over {(1-v)^{5\over 2}({3\over 2}-
                  {1\over 2}v_{max})\sqrt v_{max}}}
\nonumber \\
 &=&
 {\begin{cases}
       1.65,& \text{$m_t=0.176$ TeV},\\
       1.48,& \text{$m_t=0.199$ TeV},\\
  \end{cases}}
\end{eqnarray}
which are not inconsistent with the early CDF/D0 observations.
(For $d\bar d$ annihilation,
with $m_d(1$GeV$)\cong 8.9$MeV~\cite{leug},
the values of $r_{exp}$ are $1.74,1.56$ for $m_t=0.176,0.199$
TeV respectively.)

In summary, we have applied the soft gluon formulas in Ref.~\cite{five}
to the fundamental $t\bar t$ production process at FNAL energies.
We find that the soft gluon effects do lead to a
further significant enhancement of the
production cross section in comparision to the
usual perturbative analysis of $\sigma(t\bar t)$ at the parton level.
This enhancement is not currently inconsistent with
any known theoretical
or experimental result; indeed, it would appear to improve the
comparison between theory and experiment at this time.
We have thus developed an entirely new approach to higher order
QCD corrections in
$q\bar q \ra t\bar t +X$ based on the YFS QCD soft gluon
exponentiation formulas in Ref.~\cite{five}.
This has given us a unique prediction for
the soft multi-gluon effects in
$q\bar q \ra t\bar t +X$ and we have illustrated these effects for
the important issue of the cross section normalization at FNAL energies.
We look
forward with excitement to the further applications of our methods,
via the Monte Carlo technique~\cite{elsewhere}, to  parton-parton,
$p\bar p$ and $pp$ collisions at and above FNAL energies.

Acknowledgement:
Two of us (S. J. and B.F.L. W.) wish to thank Professors F. Gilman and
W. Bardeen of the SSCL for their kind hospitality while this work was
conceived and initiated.  These two authors also thank Prof. J. Ellis
of the CERN Theory Division for his kind hospitality while this work was
in progress.

\newpage


\end{document}